\documentclass[lettersize,journal]{IEEEtran}

\usepackage{amsmath,amsfonts}
\usepackage{algorithmic}
\usepackage{array}
\usepackage[caption=false,font=normalsize,labelfont=sf,textfont=sf]{subfig}
\usepackage{textcomp}
\usepackage{stfloats}
\usepackage{url}
\usepackage{verbatim}
\usepackage{graphicx}
\usepackage{color}
\def\BibTeX{{\rm B\kern-.05em{\sc i\kern-.025em b}\kern-.08em
    T\kern-.1667em\lower.7ex\hbox{E}\kern-.125emX}}
\usepackage{balance}
\allowdisplaybreaks[4]

\begin{document}

\title{\LARGE Geometric Constellation Shaping for Wireless Optical Intensity Channels: An Information-Theoretic Approach}

\author{Suhua Zhou, Tianqi Li, Zhaoxi Fang, Jing Zhou,~\IEEEmembership{Member,~IEEE}, and Wenyi Zhang,~\IEEEmembership{Senior Member,~IEEE}

\thanks
{This work was supported in part by Henan Key Laboratory of Visible Light Communications under Grant HKLVLC2023-B03. (\emph{Corresponding author: Jing Zhou.})

Suhua Zhou is with School of Automation, Beijing Information Science and Technology University, Beijing 100192, China (e-mail: shzhou@bistu.edu.cn).

Tianqi Li is with the National Computer System Engineering Research Institute of China, Beijing 100083, China (e-mail: litianqi@ncse.com.cn).

Zhaoxi Fang is with the Department of Computer Science and Engineering, Shaoxing University, Shaoxing 312000, China (e-mail: fangzhaoxi@usx.edu.cn).

Jing Zhou is with the Henan Key Laboratory of Visible Light Communications, Zhengzhou 450002, China, and is also with the Department of Computer Science and Engineering, Shaoxing University, Shaoxing 312000, China (e-mail: jzhou@usx.edu.cn).

Wenyi Zhang is with the Department of Electronic Engineering and Information Science, University of Science and Technology of China, Hefei 230027, China (e-mail: wenyizha@ustc.edu.cn).
}
}

\markboth{Journal of \LaTeX\ Class Files,~Vol.~18, No.~9, September~2020}
{How to Use the IEEEtran \LaTeX \ Templates}

\maketitle

\begin{abstract}
A simple geometric shaping method is proposed for optical wireless communication systems based on intensity modulation and direct detection (IM/DD) from an information-theoretic perspective.
Constellations consisting of equiprobable levels with exponential-like distribution are obtained, which possesses asymptotic optimality in the sense that the high-SNR capacity of average-intensity constrained optical intensity channel can be approached by such constellations with increasing size.
All $2^b$ levels ($b\in\mathbb N$) of the obtained constellation can be represented by a basic level and $b+2$ bits, thereby reducing the required resolution of the digital-to-analog converter (DAC) without affecting the asymptotic optimality.
Achievable information rate evaluations verify the asymptotic optimality.
As an example, error performance results of a simple $16$-level LDPC coded modulation scheme show that a shaping gain of $0.65$ dB can be obtained by applying the proposed constellation design.
This method can also be applied to more specific IM/DD channel models, since it only requires a near-optimal continuous input distribution.
\end{abstract}


\section{Introduction}
\IEEEPARstart{D}{ue} to its simplicity and low-cost in comparison with coherent systems, optical wireless communication (OWC) based on intensity modulation and direct detection (IM/DD) has been an active research area in recent years \cite{OWC}.
For example, the Li-Fi \cite{LiFi} is a promising technology for future indoor wireless networks.
In IM/DD systems, the information-bearing signal is the optical intensity (i.e., optical power per unit area).
The optical intensity channel is a widely-accepted model for indoor wireless IM/DD systems; see, e.g., studies on channel capacity [\ref{HK04}-\ref{FH10}] and transceiver design \cite{KAKK12,Shiu-Kahn}.

The focus of this letter is bandwidth-efficient transmission over wireless optical intensity channels, which require a one-dimensional multi-level constellation rather than a binary one (i.e., the on-off keying).
The bandwidth-efficient scenario is important because the transmission rate of IM/DD OWC systems is typically limited by the modulation bandwidth of light sources such as the light-emitting diode (LED) \cite{OWC}.
Since a zero-mean constellation used in conventional radio-frequency (RF) systems can be simply converted to a nonnegative one by adding a direct current (DC) bias, it is straightforward to adapt a standard pulse-amplitude modulation (PAM) constellation, which consists of $2^b$ equiprobable uniformly spaced levels ($b\in\mathbb N$), for optical intensity channels.
This also enables standard binary coded modulation techniques including capacity-achieving codes and simple bit-to-symbol mapping (labeling).

However, from an information-theoretic point of view, the standard PAM constellation is suboptimal in both RF and IM/DD systems.
In the additive white Gaussian noise (AWGN) channel, it is well-known that Gaussian input achieves capacity, and the asymptotic signal-to-noise-ratio (SNR) gap between the achievable information rate of standard PAM with unlimited number of levels and the channel capacity $C_\mathrm{AWGN}=\frac{1}{2}\log(1+\mathsf {SNR})$ (per channel use) is $1.53$ dB ($\pi e/6$) \cite{MDML20}.
That asymptotic gap is called the ultimate \emph{shaping} gain since it is the maximum possible gain of changing the shape of the input distribution from standard PAM to a Gaussian-like one.
A parallel result in the optical intensity channel is that, if we replace the standard PAM by an exponential-like constellation, then the maximum possible gain in terms of average optical power is $1.33$ dB ($e/2$) \cite{Shiu-Kahn}.
In other words, an improvement of $36\%$ in energy efficiency can be expected.

There are two basic approaches to realize the shaping gain, namely probabilistic shaping (designing non-equiprobable uniformly spaced levels) and geometric shaping (designing equiprobable non-uniformly spaced levels) \cite{MDML20,GC19}.
Different from simple equiprobable labeling for standard PAM, probabilistic shaping requires distribution matching,
which becomes more challenging as the constellation size increases.
Geometric shaping, by contrast, is easier to handle with binary channel codes when the constellation size is a power of two.
However, it results in arbitrarily spaced levels, thereby posing challenges to the resolution of digital-to-analog converter (DAC).

For the AWGN channel and RF channels, information-theoretic studies of geometric shaping can be found in, e.g., [\ref{MDML20}-\ref{SunTilborg}], in which the achievable information rate is a major performance metric, and the aim is to achieve the ultimate shaping gain (and thus achieve the capacity).
In optical communications, constellation shaping has also been investigated in an information-theoretic perspective; see, e.g., \cite{Shiu-Kahn}.
Some other studies intended to reduce error rate in the presence of various destructive issues in optical links, e.g., signal-dependent noise \cite{MTTW,Haas}, atmospheric turbulence of outdoor links \cite{JOCN22}, and intersymbol interference \cite{Kahn23}. 

In this letter, we consider the optical intensity channel and focus on the problem of approaching the channel capacity by constellation shaping.
We propose a geometric shaping method with asymptotic optimality in an information-theoretic sense,
and evaluate its performance by achievable information rate as well as error performance of coded transmission.
The idea stems from a method proposed in \cite{SunTilborg} for the AWGN channel, which leads to a Gaussian-like constellation design that achieves the ultimate shaping gain as the constellation size grows without bound.
We apply this method to the optical intensity channel to construct an exponential-like constellation, and introduce a new `shifting and scaling' step to improve the non-asymptotic performance.
Our design preserves an asymptotic property of the original method that the high-SNR capacity can be approached by increasing the constellation size.
To reduce the required resolution of the DAC, we further introduce a step of regularization, which does not affect the asymptotic property.
All $2^b$ levels of the resulting constellations can be represented by a basic level and $b+2$ bits (only two bits more than the standard PAM).
Finally we give an example through a simple $16$-level LDPC coded modulation scheme.
Numerical results show that a shaping gain of $0.65$ dB can be obtained by applying the proposed design.

\section{Optical Intensity Channel and Constellation Shaping}
Consider a discrete-time optical intensity channel with Gaussian noise as \cite{LMW09,FH10}
\begin{align}\label{OIC}
Y=X+Z,
\end{align}
where all quantities are real-valued, $X$ is the input optical intensity signal satisfying $X\ge 0$, the noise satisfies $Z\sim \mathsf N(0,\sigma^2)$, and the channel gain has been normalized to unity.
This signal-independent noise model is obtained from the classic Possion model for optical communications via several simplification procedures \cite{Chaaban}.
Our work is intended for this simplified model.
In general the variance of the noise can be signal-dependent \cite{Chaaban}.
For constellation shaping in the presence of signal-dependent noise, see, e.g., \cite{MTTW} and \cite{Haas}.

We assume an input constraint on average optical power/intensity as $\mathrm E[X]\leq \mathcal E$ \cite{LMW09,FH10}.
The channel input $X$ is drawn from a constellation $\mathcal X=\{a_0,\ldots,a_{M-1}\}$ with probability $\Pr\{X=a_i\}=p_i, \; i=0,\ldots,M-1$.
The constellation consists of $M$ real and non-negative levels (symbols) in ascending order.
Using the Dirac-delta function $\delta(\cdot)$, such an input distribution has a probability density function (PDF) as
\begin{align}\label{pX}
P_X(x)=\sum_{i=0}^{M-1}p_i\delta(x-a_i).
\end{align}
For example, a standard $2^b$-PAM constellation
satisfies $p_i=2^{-b}$ and $a_i=i\Delta$, where $\Delta$ is a basic level.

In a nutshell, for probabilistic shaping, one typically fixes a regular set of $\{a_i\}$, say, $\{i\Delta\}$, and optimize the probabilities $\{p_i\}$ to maximize the achievable information rate, while for geometric shaping, one in turn fixes an equiprobable distribution, i.e., $p_i=1/M$ for all $i$, and optimizes the levels $\{a_i\}$ to maximize the achievable information rate.
However, some practical issues should be considered in the optimization.
First, in order to apply coded modulation, there should be a labeling between binary coded bits and a level.
In probabilistic shaping it is hence customary to require every probability $p_i$ to yield a fixed-length binary expansion,
while in geometric shaping one typically lets $M=2^b$ to enable simple labeling.
Second, to reduce the required resolution of DAC, the levels of a constellation should be represented using as few bits as possible, which requires every level $a_i$ in geometric shaping to be an integer multiple of a basic level.

In the presence of an average intensity constraint, no closed-form expression for the capacity of the optical intensity channel is known, but asymptotic behaviors of the capacity at high SNR have been characterized as \cite{LMW09}
\vspace{-.1cm}
\begin{align}\label{C}
\lim\limits_{\mathsf{SNR}\to\infty}\left\{\mathsf C(\mathsf {SNR})-\frac{1}{2}\log\left(\frac{e}{2\pi}\mathsf {SNR}\right)\right\}=0,
\vspace{-.1cm}
\end{align}
where $\mathsf{SNR}:=\frac{\mathcal E}{\sigma}$ is the optical SNR defined in terms of the average optical intensity (rather than squared amplitude of the channel input).
An exponentially distributed input with PDF
\vspace{-.1cm}
\begin{align}\label{exp}
p_X(x)=\frac{1}{\mathcal E}\exp\left(-\frac{x}{\mathcal E}\right), \; x>0
\vspace{-.1cm}
\end{align}
achieves capacity asymptotically at high SNR, implying that an exponential-like constellation satisfying $\mathrm E[X]=\mathcal E$ is preferred for bandwidth-efficient transmission.

\emph{Note}:
In the AWGN channel and certain RF channels (e.g., flat-fading channel with state information at the receiver),
the capacity-achieving input distribution (CAID) is Gaussian for all SNRs.
But the CAID of the optical intensity channel with an average intensity constraint is unknown.
If there is an additional peak intensity constraint, the CAID is discrete with a finite number of mass points, which vary with the SNR \cite{CHK05}.
This letter focuses on bandwidth-efficient transmission in IM/DD systems (which require moderate to high SNR to support multi-level constellation), so that exponential distribution is near-optimal and plays a similar role as Gaussian distribution in the AWGN channel and in RF systems.

\section{Proposed Geometric Shaping Method}

\subsection{Level Generation: From Quantile to Centroid}
For a continuous probability distribution, its $M$-\emph{quantiles} are points that divide its range into $M$ equiprobable intervals.
The $M$-quantiles of the exponential PDF (\ref{exp}), denoted by $\{q_1,\ldots,q_{M-1}\}$, satisfy
\vspace{-.05cm}
\begin{align}
\frac{1}{\mathcal E}\int_{q_m}^{q_{m+1}}\exp\left(-\frac{x}{\mathcal E}\right)\mathrm d x = \frac{1}{M},\; m=0,\ldots,M-1,
\vspace{-.05cm}
\end{align}
where we let $q_0=0$ and $q_M=\infty$.
Then we obtain the $m$-th quantile as $q_m=\mathcal E\ln\frac{M}{M-m}$.
The \emph{centroid} of the interval $[q_{m},q_{m+1})$ of the exponential PDF (\ref{exp}), denoted by $c_m$, is the mean of $X$ conditioned on the event $\{q_{m}\leq X<q_{m+1}\}$.
When $0\leq m<M-1$, straightforward calculation yields
\vspace{-.1cm}
\begin{subequations}\label{cm}
\begin{align}
c_m&=\mathrm E\left[X|q_{m}\leq X<q_{m+1}\right]\\
&=\int_{q_{m}}^{q_{m+1}}x\frac{M}{\mathcal E}\exp\left(-\frac{x}{\mathcal E}\right)\mathrm d x\\
&=M\left(\left(\mathcal E+q_{m}\right)e^{-q_{m}/\mathcal E}-\left(\mathcal E+q_{m+1}\right)e^{-q_{m+1}/\mathcal E}\right)\\
&=\mathcal E\ln\frac{M}{M-m}+\varepsilon_m\\
&=q_m+\varepsilon_m,
\end{align}
\end{subequations}
where
\begin{align}
\varepsilon_m=\mathcal E\left(1-\ln\left(1+\frac{1}{M-m-1}\right)^{M-m-1}\right),
\end{align}
which satisfies $\lim_{(M-m)\to\infty}\varepsilon_m=0$.
When $m=M-1$, we have
\begin{align}\label{cm-1}
c_{M-1}=\mathrm E\left[X| X\ge q_{M-1}\right]=\mathcal E\left(\ln M+1\right).
\end{align}
Thus the centroid $c_m$ can be approximated by $\mathcal E\ln\frac{M}{M-m}$ if $M-m$ is large.
In fact, letting $M-m=k$, it can be shown that
\begin{align}
c_m=\mathcal E\left(\ln\frac{M}{k}+\frac{1}{2(k-1)}+o\left(\frac{1}{k}\right)\right),
\end{align}
where $f(k)=o(g(k))$ denotes the asymptotic relationship $\lim_{k\to\infty}\frac{f(k)}{g(k)}=0$.
We note that
\begin{align}\label{c0}
c_0=\varepsilon_0 =\mathcal E\left(1-(M-1)\ln\frac{M}{M-1}\right),
\end{align}
which goes to zero like $\frac{\mathcal E}{2(M-1)^2}$ as $M\to\infty$.

If we use the centroids to construct a constellation $\mathcal X_c=\{c_0,\ldots,c_{M-1}\}$, then we have $\mathrm E[X]=\mathcal E$ if $X$ is drawn from $\mathcal X_c$ with equal probability, thus meeting the average intensity constraint.
This can be shown by the law of iterated expectations and the fact that $c_m$ is given by a conditional expectation of $X$.
Moreover, the peak-to-average power ratio (PAPR) of this constellation is $1+\ln M$ (see (\ref{cm-1})).

In \cite{SunTilborg}, the method of using centroids of equiprobable intervals of the Gaussian PDF to construct a Gaussian-like constellation was proposed, and the asymptotic optimality of the method was proved.
In the optical intensity channel, a theoretical advantage of the constellation $\mathcal X_c=\{c_0,\ldots,c_{M-1}\}$ is given by the following two facts.
\begin{itemize}
\item
First, as $M$ increases, the achievable information rate of the constellation $\mathcal X_c$, denoted by $R_c(M)$, tends to the achievable information rate of an exponentially distributed input with mean $\mathcal E$, denoted by $I_\mathrm{exp}$:
\begin{align}\label{c2exp}
\lim\limits_{M\to\infty}R_c(M)=I_\mathrm{exp}.
\end{align}
This property can be proved following essentially the same steps as the proof of [\ref{SunTilborg}, Theorem 2].
\item
Second, as SNR increases, $I_\mathrm{exp}$ asymptotically achieves the capacity of the optical intensity channel \cite{LMW09}:
\begin{align}\label{exp2C}
\lim\limits_{\mathsf{SNR}\to\infty} I_\mathrm{exp}(\mathsf{SNR})=C(\mathsf{SNR}).
\end{align}
\end{itemize}
Thus, the high-SNR capacity of the optical intensity channel can be approached by using $\mathcal X_c$ and increasing its size.

\subsection{Level Optimization: Shifting and Scaling}
We have provided a performance guarantee for the constellation $\mathcal X_c$ in bandwidth-efficient transmission by asymptotics (\ref{c2exp}) and (\ref{exp2C}).
However, there is still room to improve its non-asymptotic performance.
Since $c_0$ is strictly positive for finite $M$ (see(\ref{c0})), we can reduce all levels by $c_0$, and compensate the reduced average intensity by a scaling factor
\begin{align}
g(M)=\frac{\mathcal E}{\mathcal E-c_0}=\left((M-1)\ln\frac{M}{M-1}\right)^{-1}.
\end{align}
This leads to new levels given by
\begin{align}\label{lm}
l_m=\frac{c_m-c_0}{(M-1)\ln\frac{M}{M-1}},
\end{align}
where $m=0,\ldots,M-1$, and yields a new constellation $\mathcal X_l=\{0,l_1,\ldots,l_{M-1}\}$.
Then the average intensity when the input is drawn from $\mathcal X_l$ with equal probability is still $\mathrm E[X]=\mathcal E$,
while the PAPR is enlarged to $1+g(M)\ln M$.
We call the preceding processing \emph{shifting and scaling}.
In fact, it is equivalent to a \emph{stretching} of the constellation which reduces the smallest level of the constellation to zero without changing its mean.

\begin{figure}[!t]
\centering
\includegraphics[width=2.4in]{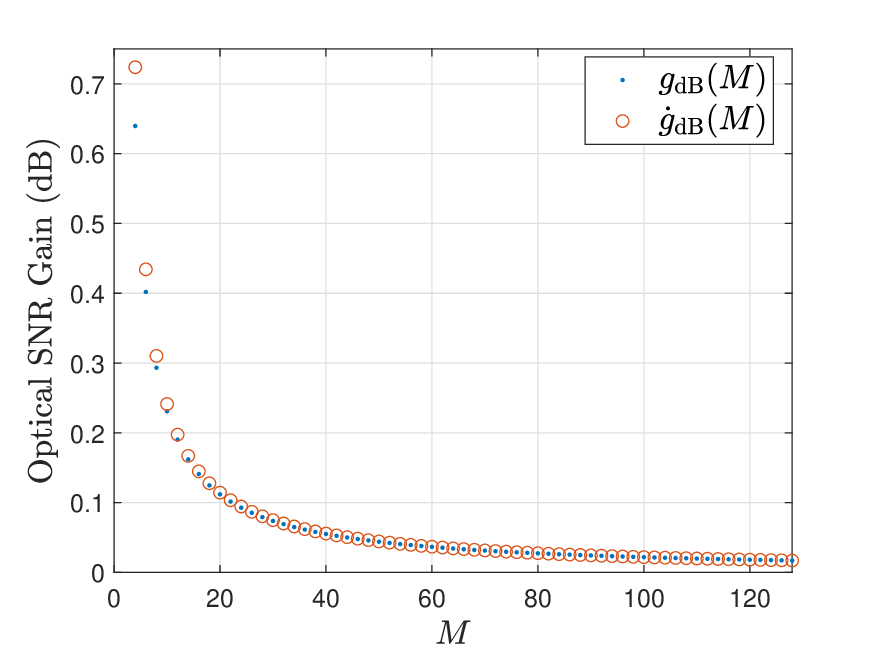}
\caption{Optical SNR gain obtained by shifting and scaling.}
\label{gg}
\end{figure}

In the presence of Gaussian noise, shifting all levels by a constant does not affect the distance between levels, and thus keeps the performance unchanged, while a scaling factor greater than one always increases the distance, and thus boosts the performance.
Therefore, for finite $M$ the constellation $\mathcal X_l$ always performs better than $\mathcal X_c$, and its gain in terms of optical SNR is exactly the scaling factor $g(M)$.
In Fig. 1, we show this gain, denoted by $g_\mathrm{dB}(M)$, for all even $M$ from 4 to 128.
In particular, when $M=8$, $16$, and $32$, the gain is $0.29$ dB, $0.14$ dB, and $0.07$ dB, respectively.
Such gains are non-negligible, given that the ultimate shaping gain is $1.33$ dB.
We also note that the gain can be approximated by
\vspace{-.1cm}
\begin{align}
\dot{g}_\mathrm{dB}(M)=\frac{5}{(M-1)\ln10}\approx\frac{2.17}{M-1},
\end{align}
i.e., it is roughly inversely proportional to $M$; see Fig. \ref{gg}.
The approximation is obtained by noting that
$g_\mathrm{dB}(M)=10\log_{10}g(M)=10\frac{\ln g(M)}{\ln 10}$
and
\vspace{-.1cm}
\begin{subequations}
\begin{align}
\ln g(M)=&-\ln\left(1-\frac{1}{2(M-1)}+o\left(\frac{1}{M-1}\right)\right)\\
=&\;\frac{1}{2(M-1)}+o\left(\frac{1}{M-1}\right).
\end{align}
\end{subequations}

\subsection{Level Regularization: Representation by $b+2$ Bits}
We have obtained a geometrically shaped constellation design $\mathcal X_l$ with closed-form level expressions and performance guarantee, where the levels can be arbitrary nonnegative real numbers.
However, in practical systems, levels are represented (with the help of a basic level $\Delta$) by finite number of bits; e.g., for standard $M$-PAM, $b=\log_2 M$ bits are enough.
We thus regularize the levels of $\mathcal X_l$ as follows.
First, let
\vspace{-.1cm}
\begin{align}\label{d}
d=\frac{l_{M-1}}{2^{b+n}-1},
\end{align}
where $n\in \mathbb N$, and map the levels of $\mathcal X_l$ to their closest points in the one-dimensional grid $\{0,d,2d,\ldots\}$ as
\vspace{-.1cm}
\begin{align}\label{ellm}
\ell_m=\left\lfloor\frac{l_m}{d}+\frac{1}{2}\right\rfloor,\;m=0,\ldots,M-1.
\end{align}
But this slightly changes the mean of the constellation.
To solve this problem, we replace $d$ by a basic level $\Delta$ determined by the relationship
\vspace{-.1cm}
\begin{align}\label{Delta}
\left(\frac{1}{M}\sum_{m=0}^{M-1}\ell_m\right)\Delta=\mathcal E.
\end{align}
The level $\Delta$ obtained in this way can be seen as a fine tuning of $d$ so that the input constraint can be perfectly matched.

Now we finally obtain a constellation consisting of $M$ real and non-negative levels as 
$\mathcal X_\mathrm{\ell}=\{0,\ell_1\Delta,\ldots,\ell_{M-1}\Delta\}$,
where $0\leq\ell_1\leq\ell_2\leq\ldots\leq\ell_{M-1}= 2^{b+n}-1$.
All the levels can be represented by the basic level $\Delta$ and $b+n$ bits.
For typical values of $b$, numerical results in Sec. IV-A will show that the performance loss is negligible when $n=2$; i.e., our geometric shaping method requires two bits more than that required by standard PAM.
In this case the obtained levels 
satisfy $0=\ell_0<\ell_1<\ldots<\ell_{M-1}= 2^{b+n}-1$; i.e., levels in $\mathcal X_l$ are mapped to different integers so that $\mathcal X_\ell$ has $M$ different levels.
As $M$ increases, the changes caused by (\ref{ellm}) become negligible so that the asymptotic properties (\ref{c2exp}) and (\ref{exp2C}) holds.

The proposed approach is summarized as follows.

{\bf{Geometric shaping procedure}}:
\begin{itemize}
\item Generate discrete levels from a continuous distribution $p_X(x)$
 by quantiles and centroids;

\item Optimize the levels by shifting and scaling;

\item Determine the final constellation by regularization and fine-tuning.
\end{itemize}

For the optical intensity channel (\ref{OIC}) considered this letter, the first step generates $\mathcal X_c=\{c_0,\ldots,c_{M-1}\}$ given by (\ref{cm}) and (\ref{cm-1}), the second step yields $\mathcal X_l=\{0,l_1,\ldots,l_{M-1}\}$ given by (\ref{lm}), and the last step determines $\mathcal X_\ell=\{0,\ell_1\Delta,\ldots,\ell_{M-1}\Delta\}$ by (\ref{d})-(\ref{Delta}).
Moreover, we note that, for a more specific IM/DD channel model than (\ref{OIC}), if a near-optimal continuous input distribution $p_X(x)$ has been found, then the above procedures can be applied straightforwardly.

\begin{figure}
\centering
\includegraphics[width=3in]{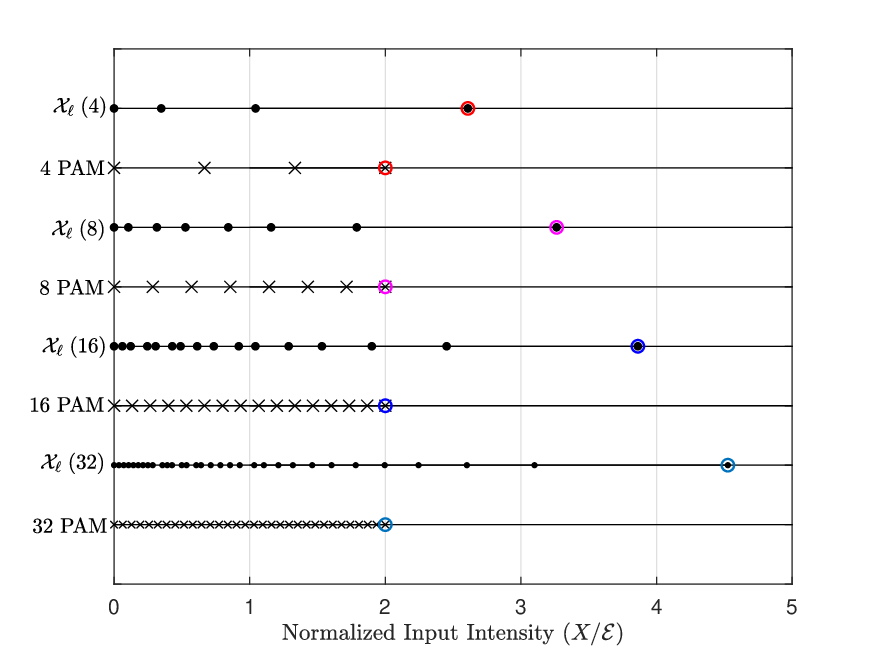}
\caption{Comparison of proposed and standard PAM constellations.
Increase of PAPR can be indicated by the largest levels (emphasized by circles).}
\label{Const}
\end{figure}

In Table I, we show the values of $\ell_m$ obtained for constellation sizes $M=4$, $8$, $16$, and $32$.
In Fig. \ref{Const}, we compare $\mathcal X_\mathrm{\ell}$ and standard PAM constellations.
The distribution of $\mathcal X_\mathrm{\ell}$ clearly exhibits exponential-like shape, and its PAPR increases roughly linearly with $b=\log_2M$ since it behaves like $\ln M$ for large $M$ (similar to that of $\mathcal X_c$ and $\mathcal X_\ell$).

\emph{Note}: Even if we only consider levels on the one-dimensional grid, finding an optimal constellation by exhaustive searching is computationally intractable for a single SNR (while the optimal constellation may vary for different SNRs).
For example, if $n=2$, we would need $\binom{2^{b+2}}{2^b}$ performance evaluations, which is approximately $5\times 10^{14}$ if $b=4$.

\begin{table}
\caption{Values of $\ell_m$ for different constellation sizes.}
\begin{tabular}{c|l}
\hline $M$ &$\{\ell_0,\ldots,\ell_{M-1}\}$\\
\hline $4$ &$\{0, 2, 6, 15\}$\\
\hline $8$ &$\{0,     1,     3,     5,    8,    11,    17,    31\}$\\
\hline $16$ &$\{0,     1,    2,     4,     5,     7,     8,    10,    12,    15,    17,    21,    25,    31,    40,    63\}$\\
\hline  &$\{0,     1,    2,     3,     4,     5,     6,     7,     8,    10,    11,    12,    14,    15,    17,    18,  20, 22$  \\
$32$ &$\mspace{8mu}24,    26,   29,   31,    34,    37,    41,    45,    50,    56,    63,  73, 87,  127\}$ \\
\hline
\end{tabular}
\vspace{-.35cm}
\end{table}

\section{Numerical Results}

\subsection{Achievable Information Rate}
In Fig. \ref{MI}, we show achievable information rates of $\mathcal X_\ell$ in the optical intensity channel (denoted by $R_\mathrm {\ell}$).
For comparison we also show achievable information rates of standard PAM (denoted by $R_\mathrm {PAM}$),
and an exponential input with mean $\mathcal E$ (denoted by $I_\mathrm {exp}$).
As a benchmark in moderate-to-high SNR region, a capacity upper bound \cite{HK04} as
$\bar{C}=\frac{1}{2}\log\left(\frac{e}{2\pi}(\mathsf {SNR}+2)^2\right)$
is also shown.
This upper bound, together with the lower bound $I_\mathrm {exp}$, determine the high-SNR capacity of the optical intensity channel as (\ref{C}).
These numerical results clearly indicate the loss of using standard PAM, and verify that the loss can be avoided by applying the proposed design.
The results are consistent with the theoretical properties given by (\ref{c2exp}) and (\ref{exp2C}).
These numerical results confirm that it is possible to approach the $1.33$ dB shaping gain in the optical intensity channel by our method.

As an example, Fig. \ref{Regular} shows the impact of the regularization step on achievable information rate of the 16-level design.
At low SNR the impact is negligible, but at high SNR we need two extra bits to avoid performance loss.
Otherwise, level regularization will \emph{combine} some constellation points, thereby considerably reducing the high-SNR limit of information rate.
\vspace{-.3cm}

\subsection{Coded Modulation Performance Results}

The information-theoretic results have confirmed the potential of our method.
In this section we provide error performance results of a simple coded modulation scheme \emph{without} any optimization regarding the non-standard constellation.
The scheme operates at moderate SNR so that the shaping gain is relatively important.
The results are obtained by MATLAB simulation with the following parameters and settings:
1) Channel code: DVB-S2 rate-$1/2$ LDPC code of length 64800 bits \cite{Sun2};
2) Constellation: $16$-level design given in Table I and Fig. 2;
3) Bit-to-symbol mapping: Gray labeling;
4) Total coding rate: $2$ bits per channel use.
5) Decoding algorithm: standard iterative decoding with at most 50 iterations.

\begin{figure}
\centering
\includegraphics[width=2.76in]{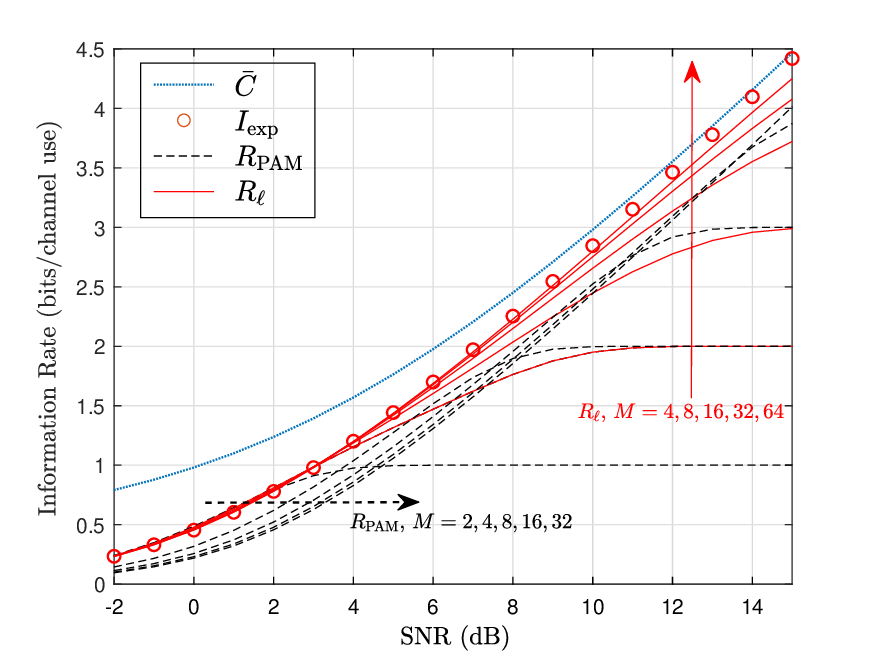}
\caption{Achievable information rate comparison.}\label{MI}
\centering
\includegraphics[width=2.76in]{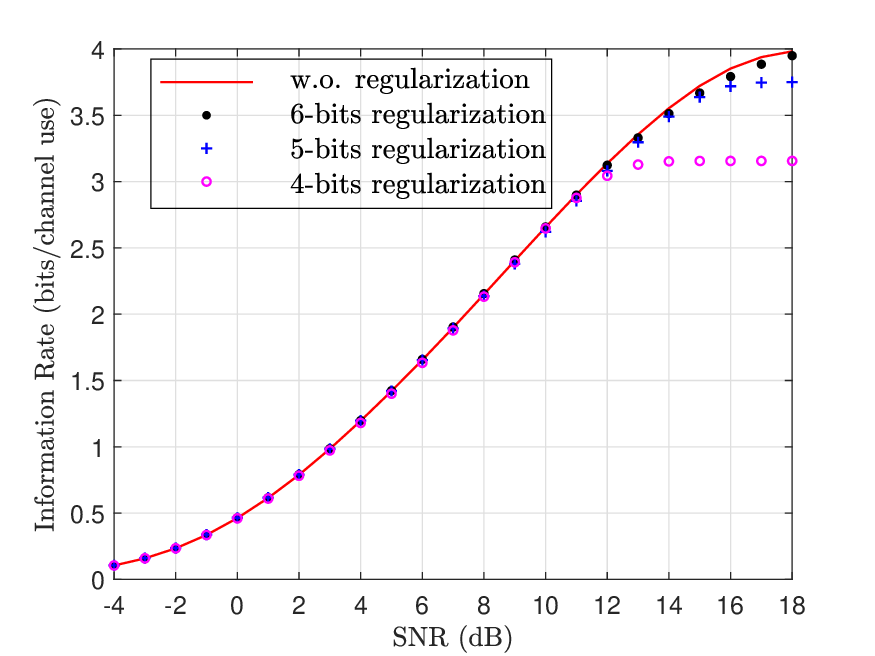}
\caption{Impact of the regularization step on achievable information rate.}\label{Regular}
\centering
\includegraphics[width=2.76in]{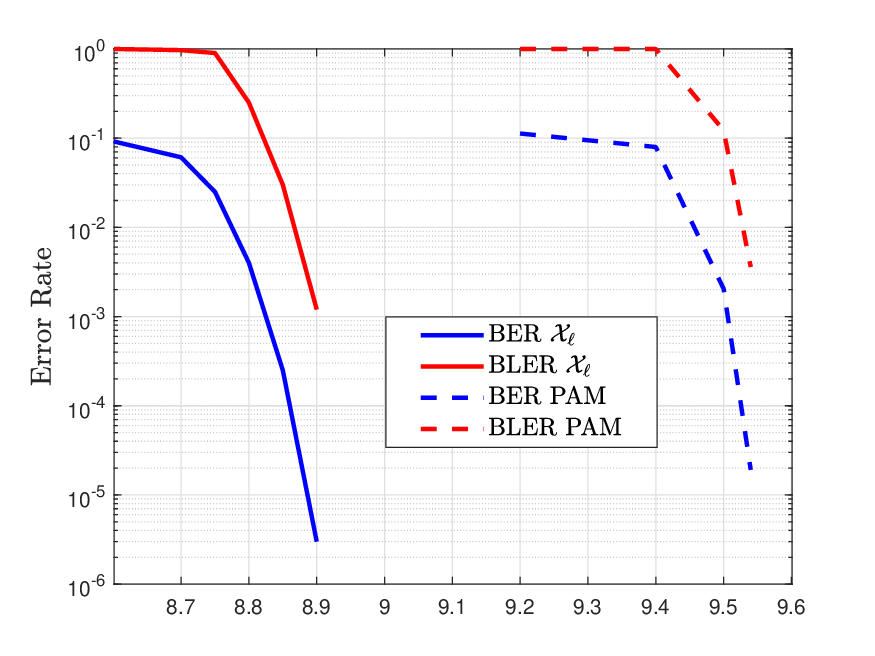}
\caption{Error performances of $16$-level LDPC coded modulation.}\label{Error}
\end{figure}

Fig. \ref{Error} shows the block error rate (BLER) and the bit error rate (BER) results of the coded modulation scheme using the proposed constellation and a standard $16$-PAM constellation.
No error floor is observed.
By simply changing the level distribution of the constellation,
we obtain a shaping gain of about $0.65$ dB (a $16$\% improvement in energy efficiency), which can be observed by comparing the SNRs required to achieve the same target error rate (e.g., a BLER of $10^{-2}$).

\emph{Note}:
In general it is highly nontrivial to approach the information-theoretic limit by a coded modulation scheme, especially when a non-standard constellation is used.
There is still a considerable gap (about 1.4 dB in optical SNR) between the performance shown in Fig. \ref{Error} and the corresponding Shannon limit (cf. Fig. \ref{MI}).
Possible causes of the gap include fundamental backoff of finite-blocklength coding, suboptimal code design and decoding (cf. \cite{Sun2}), and simple coded modulation scheme without optimization.
The problem of how to reduce the gap is left to future study.

\section{Conclusion}
We propose a geometric shaping method for bandwidth-efficient wireless optical intensity channels.
By a three-step procedure, constellations with asymptotic optimality are obtained, which require only two extra bits to represent its levels than the standard PAM.
The asymptotic optimality is confirmed by achievable information rate evaluation.
Error performance of a simple coded modulation scheme indicate that a shaping gain of about $0.65$ dB can be obtained.

\end{document}